\documentclass[pre,twocolumn,amsmath,amssymb,floats,superscriptaddress,10pt]{revtex4}

\usepackage{graphicx}
\usepackage{dcolumn}
\usepackage{bm}
\usepackage{revsymb}
\usepackage[usenames]{color}
\usepackage{subfigure}
\usepackage{color}
\usepackage{physics}
\usepackage{amsmath}
\usepackage{eufrak}
\usepackage{hyperref}

\newenvironment{psmallmatrix}
  {\left(\begin{smallmatrix}}
  {\end{smallmatrix}\right)}

\usepackage{xcolor}

\begin{document}
\title{Active viscoelasticity of odd materials}

\author{Debarghya Banerjee}
\affiliation{Max Planck Institute for Dynamics and Self-Organization, 37077 G\"{o}ttingen, Germany}
\affiliation{Max Planck Institute for the Physics of Complex Systems, 01187 Dresden, Germany}
\author{Vincenzo Vitelli}
\affiliation{James Franck Institute, The University of Chicago, Chicago, Illinois 60637, USA}
\affiliation{Department of Physics, The University of Chicago, Chicago, Illinois 60637, USA}
\affiliation{Kadanoff Center for Theoretical Physics, The University of Chicago, Chicago, Illinois 60637, USA}
\author{Frank J\"{u}licher}
\affiliation{Max Planck Institute for the Physics of Complex Systems, 01187 Dresden, Germany}
\affiliation{Cluster of Excellence Physics of Life, TU Dresden, 01062 Dresden, Germany}
\author{Piotr Sur\'{o}wka}
\email{surowka@pks.mpg.de}
\affiliation{Max Planck Institute for the Physics of Complex Systems, 01187 Dresden, Germany}
\affiliation{W\"{u}rzburg-Dresden Cluster of Excellence ct.qmat, Germany}
\affiliation{Department of Theoretical Physics, Wroc\l{}aw  University  of  Science  and  Technology,  50-370  Wroc\l{}aw,  Poland}

\begin{abstract}
The mechanical response of active media ranging from biological gels to
living tissues is governed by a subtle interplay between viscosity and elasticity. In this Letter, we generalize the canonical Kelvin-Voigt and Maxwell models to active viscoelastic media that break both parity and time-reversal symmetries. The resulting continuum theories exhibit viscous and elastic tensors that are both antisymmetric, or odd, under exchange of pairs of indices. We analyze how these parity violating viscoelastic coefficients determine the relaxation mechanisms and wave-propagation properties of odd materials.

\end{abstract}

\maketitle

Materials at the macroscopic scales are often described either
as a fluid or as a solid. Such idealized behaviors are insufficient to describe materials that exhibit more complex mesoscopic organization. This is for example the case for liquid crystals or gels, where the structure of matter is more intricate due to
elongation or chirality of the constituents. In addition, dissipative or active processes may enter the description of microscopic building blocks. As a result, the macroscopic
description of materials includes both fluid- and solid-like features -- the interplay of these two elements is known as
viscoelasticity (see e.g. \cite{lakes2010}).  Viscoelasticity is a common
phenomenon, described by rheology, that can be observed in polymer systems \cite{ward2004}, metamaterials \cite{Bertoldi2017,Shankar2020} and various
biological media \cite{Kruse2004,Frthauer2012}.  

From the swimming strokes of sperm cells to intracellular flows -- biological systems present a wide
variety of cases where chiral symmetry is broken \cite{Tsai2005,Henley2009,Vandenberg2009,Henley2012,Frthauer2013,Naganathan2014}. Additionally, biological
systems are often driven away from thermodynamic equilibrium by chemical
reactions that render the matter \emph{active} at the molecular level
\cite{Julicher2018,Marchetti2013,Ramaswamy2010}. 
Recent work on \emph{chiral} active matter has shown that the presence of
activity and chirality, breaking essential microscopic symmetries, leads to
novel response functions and transport coefficients in active fluids and solids
\cite{Banerjee2017,Salbreux2017,Scheibner2019,MaitraLenz2019,MaitraRamaswamy2019,
Hoffmann2020,Sumino2012,Tabe2003,Petroff2015,Riedel2005,Denk2016,Lenz2003,Uchida2010,vanZuiden2016}.
In the simplest incarnation, the two coefficients were
dubbed odd viscosity \cite{Avron1995,Avron1998,Lapa2014,Lucas2014,Kogan,Ganeshan2017,Liao2019,Souslov2019,SouslovPRL2019,Soni2019}
and odd elasticity \cite{Scheibner2019}. The main goal of the present work is to combine these two formulations through a systematic description of chiral active systems based on symmetry principles. This leads to a hydrodynamic theory of active odd viscoelastic solids and liquids, distinguished by a long-time response to static and dynamic deformations. Odd responses are absent in standard Kelvin-Voigt and Maxwell models.

{\it Basic viscoelastic models}.$-$Viscoelasticity emerges as a consequence of complex phenomena at different
scales. It is usually not possible to have a first principle analysis and
various phenomenological simplifications are employed. Since viscoelasticity can be viewed as a transient phenomenon to a time-independent
state we need to make some assumptions about the long time behavior of our viscoelastic
system. It can be either fluid or solid. These two distinct limiting cases of viscoelastic behavior are commonly described by the Maxwell and Kelvin-Voigt models respectively. These
models can be presented by a spring and a viscous damper in series (for Maxwell
materials) or in parallel (for Kelvin-Voigt materials). The Kelvin-Voigt model
typically defines a viscoelastic solid and captures strain relaxation. Maxwell materials define viscoelastic fluids, representing stress relaxation. 
General viscoelastic response in rheology is defined for small deformations as
\begin{equation} \label{eq:memory}
\sigma_{ij}(t) = \int _{-\infty}^{t} \eta_{ijkl}(t-\tau)\frac{d u_{kl}}{d \tau}d\tau,
\end{equation}
where $\sigma_{ij}$ is the stress tensor, $u_{kl}$ corresponds to strains. $\eta_{ijkl}(t)$ is the generalization of the elasticity tensor to viscoelastic systems with memory kernel. It is crucial to keep the tensor structure, when symmetries are broken. Causality requires that $\eta _{ijkl}(t)$ is zero for negative $t$. $\eta _{ijkl}(t)$ captures the response to a constant shear rate applied to the material, in such a way that symmetries are preserved. We always imply a summation over repeated indices, which run over spatial $x$ and $y$ coordinates as we focus on two-dimensional systems. Solids and liquids differ by the limiting value of the stress under a deformation, which, at long times, tends to zero for fluids and to a constant value for solids.
As an equivalent description of viscoelastic response one can consider the stress evolution as an input using the creep compliance tensor $c_{ijkl}(t)$ in the following way \cite{pipkin1986lectures}
\begin{equation}\label{eq:creep}
u_{ij}(t)= \int _{-\infty}^{t} c_{ijkl}(t-\tau)\frac{d \sigma_{kl}}{d \tau}d\tau.
\end{equation}
Creep is a progressive deformation of a material under constant stress. If it approaches a finite shear, the material is said to be solid if it increases linearly after a long time, the material is a fluid.

We are interested in linear viscoelasticity and focus on the simple case, where the stress at the current time depends only on the current strain and strain rate. As a simple illustrative example, we note that, for the Kelvin-Voigt model of a solid, the relation between strain and stress is given by
\begin{align}\label{eq:KV4}
\sigma_{ij} = \kappa_{ijkl} u_{kl} + \eta_{ijmn} {\dot u}_{mn},
\end{align}
where $\sigma_{ij}$ is the stress tensor, $u_{ij}$ is the strain tensor,
${\dot u}_{ij}$ is the strain rate tensor, $\kappa_{ijkl}$ is the elasticity
tensor, and $\eta_{ijkl}$ is the viscosity tensor. Here, $\sigma_{ij}$ and $u_{ij}$ are symmetric and $\kappa_{ijkl}$ and $\eta_{ijmn}$ are symmetric in the first two and last two indices. Considering linear viscoelasticity and isotropy reduces the number of entries in $\kappa_{ijkl}$ and $\eta_{ijmn}$ to two in the parity even case. These are the first and second Lam\'{e} parameters $\lambda$ and $\mu$ as well as two viscosities, bulk viscosity $\zeta$ and shear viscosity $\eta$.
This approach was first proposed by Maxwell in his spring and damper model and later complemented by Kelvin, Voigt and also Meyer \cite{Nhan} (see also \cite{Azeyanagi2009,Fukuma2011,Armas2020} for modern developments). 

In this Letter, following the same logic, we employ linearized viscoelasticity to construct parity-breaking generalized Kelvin-Voigt and Maxwell models in two dimensions that also break time-reversal symmetry, which we use to investigate the parity odd viscoelastic response in solids and fluids. In the present case the parity breaking manifests itself through a non-zero value of the odd viscosity $\eta^o$ and the odd elastic coefficient $\kappa^o$. Odd elasticity encapsulates non-conservative microscopic interactions.  This means that in a cyclic process, the net elastic work $\oint \sigma_{ij} du_{ij} $ can be nonzero in the presence of odd elasticity and the sign of the net elastic work changes when the cyclic process is reversed. Therefore odd elasticity \cite{Scheibner2019} breaks time reversal symmetry and only exists in active systems \cite{Salbreux2017,Scheibner2019}. Note that odd viscosity requires parity breaking but can exist in a passive system as it is consistent with time reversibility \cite{Hoyos:2014pba}.

\begin{figure}
\centering
\includegraphics[width= \linewidth]{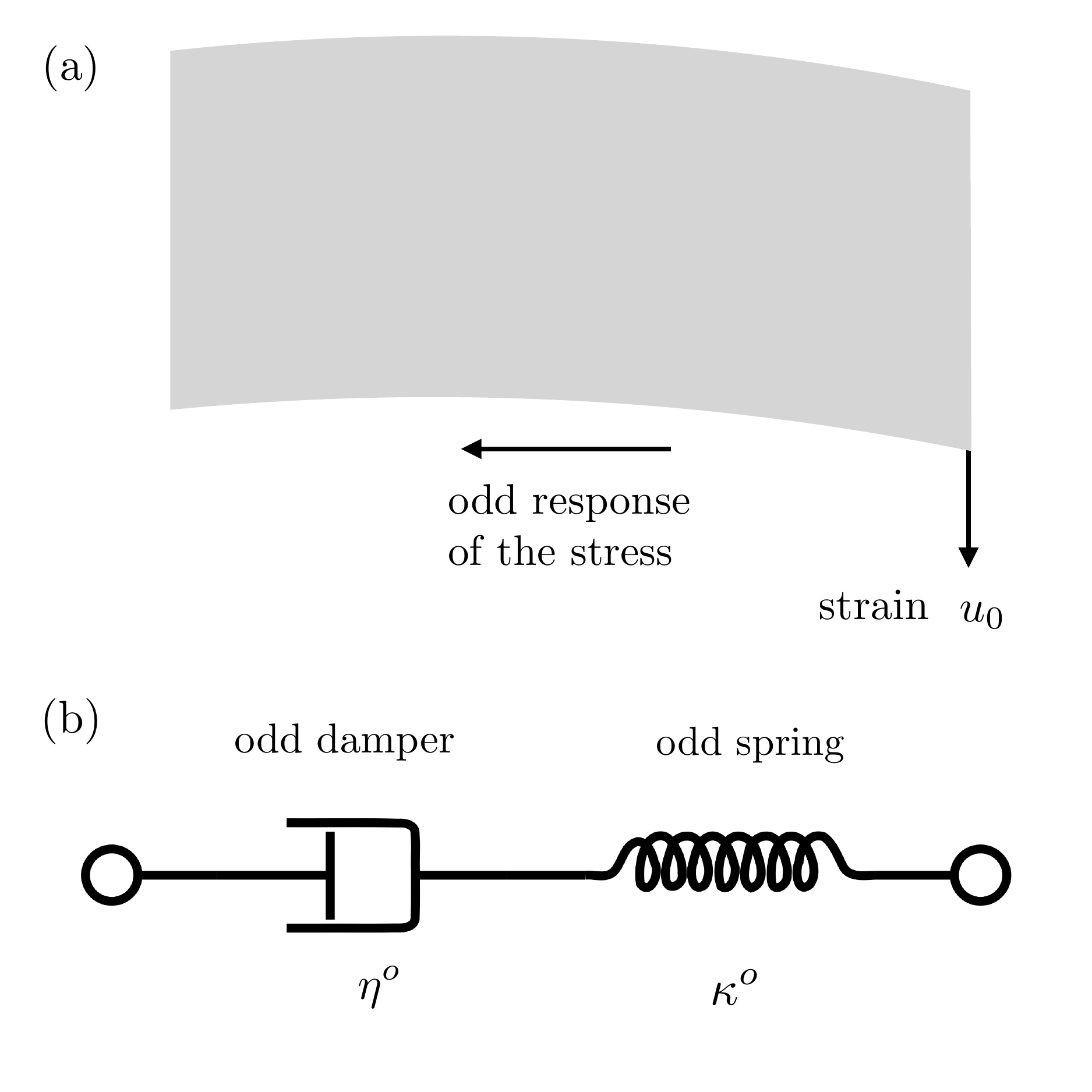}
\caption{(a) A cartoon of the odd Maxwell model. A finite deformation will result in a transverse stress. Its relaxation will be captured by odd transport coefficients. (b) Depiction of the odd response
in odd Maxwell materials. It is modelled by a spring and a damper connected in series. We emphasize that this collective representation \emph{does not} imply that odd Maxwell fluids can be constructed out of springs and dampers.}\label{fig:Max}
\end{figure}

{\it Relaxation of parity-odd viscoelastic materials in two dimenions}.$-$The relaxation times of a viscoelastic 
system tell us how long it takes for the material to return from a deformed state to its equilibrium state. In two dimensions there are two
distinct types of stresses one can apply to the material: shear and compression. 
As we shall see, parity breaking does not modify the compressional response. However, 
the response to shear receives contributions from odd transport coefficients. 
We are interested in determining the explicit form of the relaxation times. For the Kelvin-Voigt model we rewrite the constitutive equation in the following form
\begin{equation}\label{eq:relax2}
 R_{ijkl}\dot{u}_{kl} = -u_{ij}+\kappa _{ijkl}^{-1}\sigma_{kl},
\end{equation}
where $R_{ijkl} \equiv \kappa_{ijmn}^{-1}\eta_{mnkl}$ and $i,j,k,l=\{1,2\}$, which we refer to as the relaxation times tensor. In the Kelvin Voigt model $u_{ij}$ corresponds to the deformation from a specific reference state. Analogously for the Maxwell model in the presence of a flow given by the velocity $\vec{ v}$ the constitutive relation reads
\begin{equation}\
R_{ijkl}  \frac{d}{d t} \sigma_{kl}= -\sigma_{ij}+ \eta_{ijkl} v_{kl}.
\end{equation}
where, $d/d t =\partial _t +v_i \partial _{i}$ is the convective derivative and $v_{kl}=\frac{1}{2} (\partial_k v_l + \partial_l v_k)$ is the symmetric part of the velocity gradients. In general the corotational part should also be included \cite{Julicher2018} but we omit it for simplicity. Four-index tensors possess a unique identity operator $
\text{Id}_{ijkl}= \delta_ {ik}\delta_{jl}$, where $\delta_{ij}$ denotes the Kronecker symbol.
In order to define $\kappa^{-1}$ we need to construct a tensor that contracted with the elasticity tensor 
$\kappa_{mnkl}$ satisfies $
\kappa ^{-1}_{ijmn}\kappa_{mnkl} =\text{Id}_{ijkl}.
$
\begin{figure}
\centering
\includegraphics[width= \linewidth]{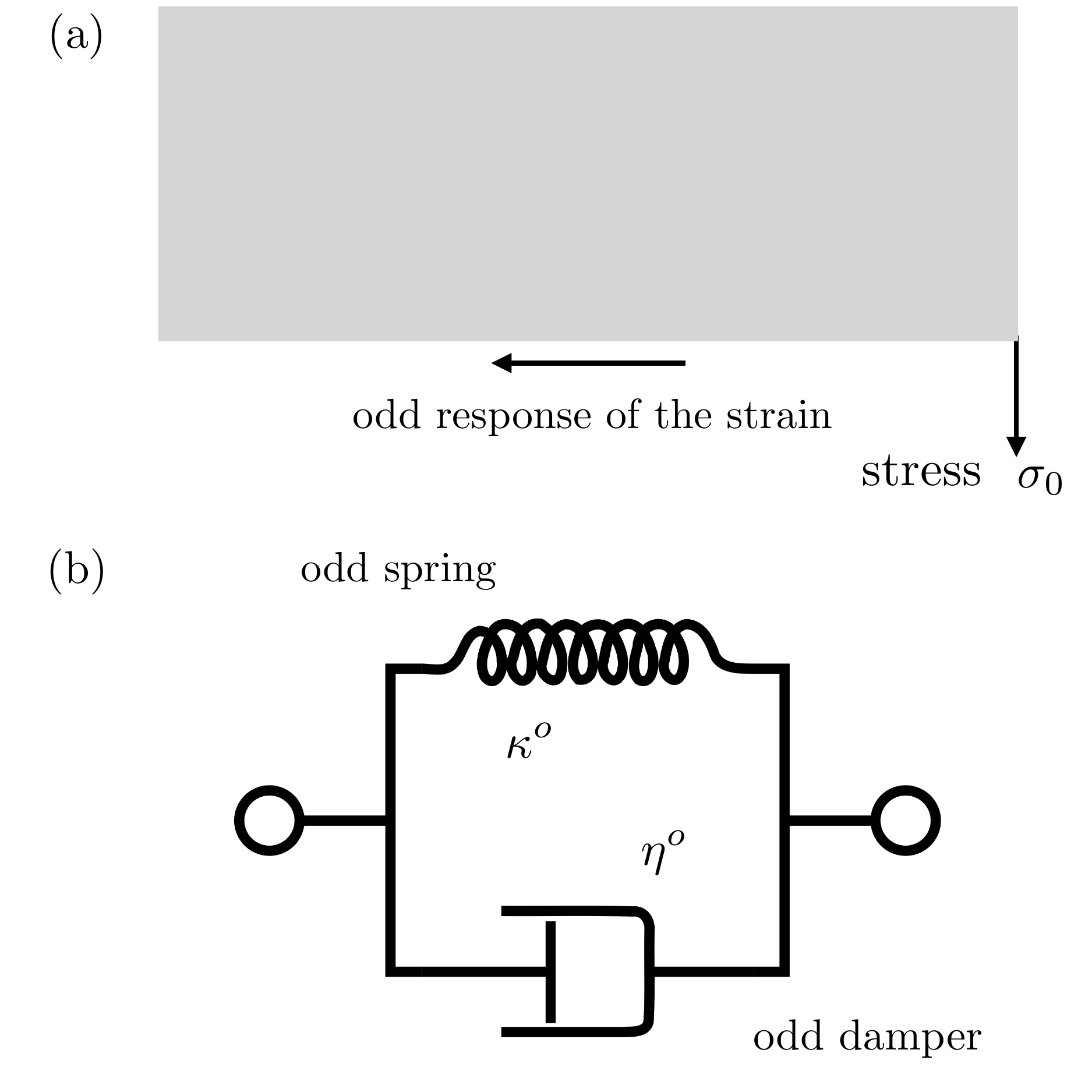}
\caption{(a) A sketch of the odd Kelvin-Voigt model. Upon application of a constant stress the resulting deformation will be perpendicular to the stress direction. The decay of the deformation will be captured by odd transport coefficients. (b) Depiction of the odd response in odd Kelvin-Voigt materials. It is modelled by a transverse spring and a damper connected in parallel.} 
\label{fig:KV}
\end{figure}
We note that in classical elasticity the elasticity tensor is only partially invertible and we construct the inverse in the invertible subspace. 
Assuming that there is no memory in the system, we can use the symmetries to 
extract the structure of the relaxation-times tensor for an isotropic two-dimensional material
\begin{align}\label{eq:relax}
 R_{ijkl} &= \tau_1 (\delta_{ik} \delta_{jl} + \delta_{il} \delta_{jk}) + \tau_2 \delta_{kl} \delta_{ij} \\  \nonumber
 &+ \frac{\tau^o}{2} \left( \delta_{ik} \epsilon_{jl} + \delta_{il} \epsilon_{jk} + \delta_{jk}\epsilon_{il} + \delta_{jl}\epsilon_{ik} \right)  \nonumber
\end{align}
in terms of the three coefficients $\{\tau_1,\tau_2,\tau^o\}$. Here $\epsilon_{ij}=-\epsilon_{ji}$ denotes the antisymmetric tensor with odd parity.
On symmetry grounds the system exhibits three relaxation times. 
These times $\tau_1=\frac{\eta^o \kappa^o+\eta \mu}{4((\kappa ^o)^2+\mu ^2)}$, $\tau_2=\frac{\zeta +\eta}{4(\lambda +\mu)}-\tau_1$, $\tau^o=\frac{\eta ^o \mu-\eta \kappa^o}{4((\kappa ^o)^2+\mu ^2)}$ depend on the transport coefficients $\zeta$, $\lambda$, $\eta$, and $\mu$ defined below. $\tau _1$ and $\tau_2$ capture the relaxation of compression and shear respectively. The relaxation time $\tau^o$ is associated with the relaxation of perpendicular responses as depicted in Fig. \ref{fig:KV} for the Kelvin-Voigt solid and in Fig. \ref{fig:Max} for the Maxwell fluid. We note, however, that in general all these relaxation processes are coupled leading to an oscillatory relaxation dynamics.

In order to get a better physical understanding of the symmetry structure, it is convenient to use a basis of 
two-dimensional matrices
\begin{equation}\label{eq:2dk}
    s^0 = \mqty(1 & 0 \\ 0 & 1), \quad s^2 = \mqty(1 & 0 \\ 0 & -1 ),\quad s^3 = \mqty(0 & 1 \\ 1 & 0).
\end{equation}
In this basis the stress and displacements are defined as $\sigma^\alpha = s^\alpha _{ij} \sigma_{ij}$, $u^\alpha=s^\alpha _{ij} u_{ij}$.
where $\alpha =\{0,2,3 \}$, see Ref. \cite{Scheibner2019} for details. The absence of $s^1= \begin{psmallmatrix}0 & -1\\1 & 0\end{psmallmatrix}$ stems from the assumption that stress and deformation tensors are symmetric. In this representation the elastic tensor is given by a matrix with elements $\kappa ^{\alpha \beta} =\frac{1}{2} (s^\beta)^{-1}_{ij} \kappa_{ijmn} s^\alpha_{mn}$. The original form can be obtained using the formula $ \kappa_{ijmn}=\frac{1}{2}(s^\beta)_{ij}\kappa ^{\alpha \beta}  (s^\alpha)^{-1}_{mn}$.
In order to see the explicit form of the elastic tensors in these two bases we start with the most general form of the elastic tensor 
\begin{align}\label{eq:coeffs}
 \kappa_{ijkl}& =  \mu (\delta_{ik} \delta_{jl} + \delta_{il} \delta_{jk}) + \lambda \delta_{kl} \delta_{ij} \\ \nonumber
 &+ \frac{\kappa^o}{2} \left( \delta_{ik} \epsilon_{jl} + \delta_{il} \epsilon_{jk} + \delta_{jk}\epsilon_{il} + \delta_{jl}\epsilon_{ik} \right),
\end{align}
and the viscosity tensor 
\begin{align}\label{eq:coeffs}
 \eta_{ijkl} &= \eta (\delta_{ik} \delta_{jl} + \delta_{il} \delta_{jk}) + \zeta \delta_{kl} \delta_{ij} \\ \nonumber
 &+ \frac{\eta^o}{2} \left( \delta_{ik} \epsilon_{jl} + \delta_{il} \epsilon_{jk} + \delta_{jk}\epsilon_{il} + \delta_{jl}\epsilon_{ik} \right),
\end{align}
consistent with  spatial isotropy and the minor symmetry of indices. Using the representation in Eq. \eqref{eq:2dk} we obtain
\begin{align}
       \kappa^{\alpha \beta}= 2\mqty(
\lambda +\mu & 0 & 0 \\  
 0 & \mu & -\kappa ^o \\
 0 & \kappa ^o & \mu \\
)^{\alpha \beta}, \label{isoform}
    \end{align}
and    
   \begin{align}
       \eta^{\alpha \beta}= 2\mqty(
\zeta +\eta & 0 & 0 \\  
 0 & \eta & -\eta ^o \\
 0 & \eta ^o & \eta \\
)^{\alpha \beta}. \label{isoform}
    \end{align} 
Now the inversion requires an inverse of a $3\times3$ matrix.
The relaxation tensor in Eq. \eqref{eq:relax} has a simple form
$
 R_{\alpha \beta} =\frac{1}{2} \kappa_{\alpha \gamma}^{-1}\eta_{\gamma \beta}.
$
It is also convenient to introduce an inverse of the relaxation times tensor \begin{equation}\label{eq:invrelax}
 \Lambda_{ijkl} = \eta_{ijmn}^{-1}\kappa_{mnkl},
 \end{equation}
that we call relaxation rate tensor, whose matrix representation is
$
 \Lambda_{\alpha \beta} = \frac{1}{2}\eta_{\alpha \gamma}^{-1}\kappa_{\gamma \alpha}= R_{\alpha \beta} ^{-1} .
$

{\it Modified Kelvin-Voigt material}.$-$The first viscoelastic material we want
to investigate is an active, chiral, viscoelastic solid (see Figure \ref{fig:KV}). 
Such a material shall
have parity breaking viscous and elastic response transport coefficients together with the conventional parity preserving responses.
It has been shown that odd elastic solids require some dissipation mechanism to
make them stable \cite{Scheibner2019}. Such a mechanism can be provided by a small viscosity. This
type of viscoelastic solid was studied in \cite{Scheibner2019}. In the present
study we want to focus on the interplay between odd elastic and odd viscous
effects. To do that one can extend the Kelvin-Voigt model to account for parity
breaking responses. We start with the constitutive equation \eqref{eq:KV4} written in the matrix representation:
\begin{align} \label{eq:KV}
\sigma_{\alpha} = \kappa_{\alpha \beta} u_{\beta} + \eta_{\alpha \gamma} {\dot u}_{\gamma}.
\end{align}
In this model the stresses due to viscosity and elasticity are additive and they can be represented by a spring and a
damper in parallel. Odd elastic solids have been modelled by using point masses connected by springs that have linear but non-central forces \cite{Scheibner2019, Zhou2019, Scheibner2020}, the odd Kelvin-Voigt model is the simplest extension of such solids to include damping or viscous effects.

The resulting odd response in the continuum limit is perpendicular to the applied force.
In the case of viscoelastic solids modelled by the Kelvin-Voigt equations \eqref{eq:KV} we are interested in the relaxation of displacement. To understand this we have to analyze the eigenvalues of the relaxation time tensor.  
We get one time that corresponds to compression that is real and does not depend on parity breaking coefficients
\begin{equation} \label{eq:relaxtcompress}
\tilde{\tau}=\frac{1}{2}\frac{\eta +\zeta}{\mu+\lambda},
\end{equation}
and two complex times that determine the relaxation of the deviatoric perturbations
\begin{equation} \label{eq:relaxtshear}
\tilde{\tau}_{1,2}=\frac{1}{2}\frac{\eta ^o \kappa^o+\eta \mu \pm i (\eta ^o \mu -\eta \kappa ^o)}{\mu^2 +(\kappa ^o)^2}.
\end{equation}
We conclude that in the parity breaking viscoelastic solids shear perturbations create damped oscillating waves noticed in \cite{Scheibner2019} without odd or Hall viscosity. We also note that in the limit with $\kappa ^o=0$ the waves still exist and the imaginary part is given by the product of odd viscosity and an even elastic coefficient $\mu$.

{\it Modified Maxwell material}.$-$Soft materials like dilute biopolymer solutions can also exhibit viscoelastic
behavior \cite{Prost2015,FischerFriedrich2016,Pegoraro2017,Saintillan2018}. They can transiently store elastic energy but they can also flow like viscous
fluids.

We would like to have a model that can serve as a description of chiral active 
polymer solutions with odd viscoelastic terms. 
We follow a phenomenological path by taking a Maxwell model, in which $\sigma_{ij} = \sigma^D_{ij} = \sigma^S_{ij},$ $u_{kl} = u^D_{kl} + u^S_{kl}$, and supplement it with parity breaking terms.
The superscript $D$ and $S$ refer to damper and spring respectively. 
In the Maxwell model
the strain is additive and modelled by a spring and a damper 
in series (as opposed to being parallel in the Kelvin-Voigt model) as presented in Figure \ref{fig:Max}.
The constitutive equation together with the momentum conservation equation for the Maxwell fluid read (see also Fig. \ref{fig:Max}):
\begin{subequations}
\begin{align}\label{eq:Maxfl}
& v_{kl} =\eta_{ijkl}^{-1} \sigma_{ij} + \kappa_{ijkl}^{-1} \frac{d}{d t} \sigma_{ij},  \\
&\rho \frac{d}{d t}  v_i=  \partial _j \sigma _{ij},
\end{align}
\end{subequations}
where $\rho$ is the mass density. We can rewrite the above equations using the inverse of the relaxation times tensor \eqref{eq:invrelax}.
This model together with the relaxation times analysis for broken parity and time reversal symmetries is a central result of our study. We perform analytic studies in a simplifying limit. 
We consider an analytically tractable limit, taking small elastic coefficients $\kappa_{ijkl}$ , while keeping the relaxation rates $\Lambda _{ijmn}$ fixed. 
The relaxation times are given again by the expressions \eqref{eq:relaxtcompress} and \eqref{eq:relaxtshear}. In a complete analogy with Kelvin-Voigt material we see that changing the volume of the material relaxes without any influence from activity. On the other hand area preserving shear deformations lead to oscillating, damped shear waves.

{\it Compressible Maxwell fluids}.$-$Parity breaking affects pressure in a significant way \cite{Lucas2014}. We can investigate this in the compressible regime of the odd
Maxwell model. Let us now look at the equations \eqref{eq:Maxfl} linearized around a solution with no velocity and $\rho =\rho_0$ governing the dynamics
of an odd viscoelastic Maxwell fluid:
\begin{subequations}
\begin{align}
 \frac{\partial}{\partial t} \sigma_{ij} & = \mu \left( \partial_i v_j + \partial_j v_i \right) + \lambda \delta_{ij} \partial _k v_k\\ & +\frac{\kappa^o}{4} \left( \partial_i^* v_j + \partial_i v_j^* + \partial_j^* v_i + \partial_j v_i^* \right) - \Lambda_{ijmn} \sigma_{mn} ,\nonumber \\
 \rho_0 \frac{\partial}{\partial t} v_i & = \partial_j \sigma_{ij}, \\
 \frac{\partial}{\partial t}\delta \rho & = -\rho_0 \partial _k v_k,
\end{align}
\end{subequations}
where $\partial_i^*=\epsilon_{ij}\partial _{j}$ and $v_i^*= \epsilon_{ij}v_j$, $\rho=\rho_0 +\delta \rho$, and $p =-\frac{1}{2} \sigma _i^{{\,\,}i}$ (see also \cite{Bollada2012}). In order to make analytic progress we take a limit $\eta^o/\eta \rightarrow 0$ and $\zeta/\eta\rightarrow 0$ simultaneously. Next we rewrite the resulting equations using vorticity $\Omega=\epsilon_{ij} \partial _i v_j$ and the divergence $\Theta = \partial _k v_k$ as variables
\begin{subequations}
\begin{align}
&\rho _0 \frac{\partial^2}{\partial t^2} {\Omega} = \mu \nabla^2 {\Omega} - \kappa^o \nabla^2 {\Theta}, \\
&\rho _0 \frac{\partial^2}{\partial t^2} {\Theta} =  \left( 2 \mu + \lambda \right) \nabla^2 {\Theta} + \kappa^o \nabla^2 {\Omega}.
\end{align}
\end{subequations}

\begin{figure}[t!]
\includegraphics[width=\linewidth]{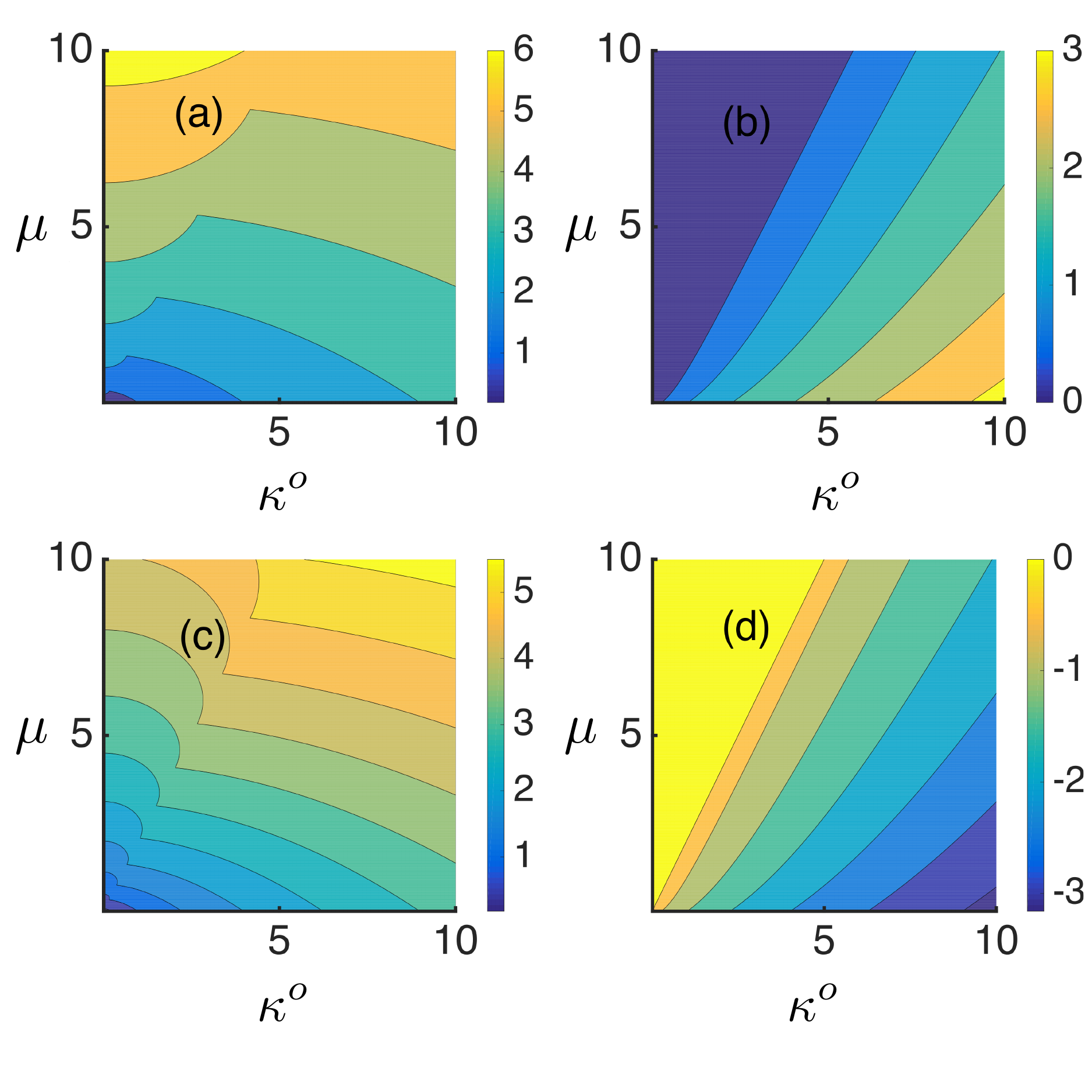}
\caption{Real and imaginary parts of the frequency eigenvalues in (a) and (b) we plot the real and imaginary parts respectively, of the first eigenvalue; while in (c) and (d) we plot the real and imaginary part of the second eigenvalue respectively. These are obtained for the parameters $k = 1$, $\lambda= 0$, and $\rho_0 = 1/2$.} \label{fig:eig} 
\end{figure}
These equations allow vorticity and divergence sound waves. Now we study plain-wave perturbations $\Omega (t,\vec{x})=\Omega (\omega,\vec{k}) e^{i(\omega t+\vec{k}\cdot \vec{x})}$ and $\Theta (t,\vec{x})=\Theta (\omega ,\vec{k}) e^{i(\omega t+\vec{k}\cdot \vec{x})}$. In the usual viscoelastic theory, longitudinal waves corresponding to the compressional disturbances are decoupled from the transverse waves that correspond to shear perturbations \cite{rogerborcherdt2009}. Odd viscoelasticity couples these two collective disturbances. These linear waves have dispersion relations given by:
 \begin{align}
&\omega = \pm k \frac{1}{\sqrt{2\rho_0}}  \left[\left(\lambda+3\mu \right) 
\pm \left((\lambda +\mu)^2 - 4{\kappa^o}^2 \right)^{1/2} \right]^{1/2}. 
\end{align} 
\color{black}{
We obtain two modes for each equation. The eigenvalues are plotted in Fig. \ref{fig:eig}. Now, considering the extreme case of $\kappa^o \gg \lambda,\mu$ we have:
\begin{align}
\omega = \pm k \frac{1}{\sqrt{\rho_0}}  \sqrt{\kappa^o} \left( \pm i \right)^{1/2}
\end{align}
The above dispersion relation yields waves that have speeds proportional to $\sqrt{\kappa^o}$ and also
linear damping proportional to $\sqrt{\kappa^o}$ in the stable case with $\kappa^o>0$. They are distinct from Avron waves in fluids with odd viscosity \cite{Avron1998}, which are of diffusive nature. As such they provide a novel manifestation of parity breaking excitations in hydrodynamics.

{\it Discussion}.$-$
We have focussed our study of odd viscoelastic responses, in chiral active matter, on two-dimensional systems. Our motivation comes from a particular pertinence of chiral asymmetries near surfaces as in two dimensions. An important example of such an active chiral
system in biology is the cell cortex. This is a thin layer of an active gel formed by actin filaments and
many other protein components which is located below the cell membrane. This gel layer has viscoelastic
properties because filaments turn over in about one minute giving its elastic properties on short time scales
and viscous behaviors at long times \cite{FischerFriedrich2016}.
It has been shown that this system generates active stresses in the layer and active chiral processes give rise to asymmetric flows \cite{Salbreux:2009p17477,Bois:2011kx,Gross:2018bg}.
Other examples of two-dimensional viscoelastic systems include
interfaces between complex active fluids, thin layers of viscoelastic substances placed between plates, and solid meta-materials constructed with active/non-reciprocal microscopic interactions, as described in \cite{Scheibner2020}, where the Kelvin-Voigt model of viscoelasticity is directly applicable.

Odd active viscoelasticity can, in principle, be tested in microrheological experiments, where viscoelastic properties of complex fluids can be determined from the motion of embedded colloidal particles \cite{SquiresMason}. 
Based on symmetry arguments, the dynamics of a tracer particle in a odd viscoelastic medium would be described by an equation of the form
\begin{equation}\label{eq:Bead}
m \dot{V}_i (t) = F_i  - \int _0 ^t dt'\begin{psmallmatrix} \eta(t-t') & -\eta^o(t-t')\\ \eta^o(t-t') & \eta(t-t')\end{psmallmatrix}_{ij} V_j (t')  .
\end{equation}
Here $\vec{V}$ is the velocity of the bead of mass $m$, $\vec{F}$ denotes an applied force and the memory kernels $\eta(t)$ and $\eta^o(t)$ capture
normal and odd viscoelastic responses respectively.
In a parity breaking viscoelastic fluid the drift of a particle should be governed by odd transport coefficients. This suggests that signatures of effects associated with odd viscoelasticity could be observed in the dynamics of particles embedded in chiral active gels.

{\it Acknowledgements}.$-$PS was supported by the Deutsche Forschungsgemeinschaft through the Leibniz Program, the cluster of excellence ct.qmat (EXC 2147, project-id 39085490) and the National Science Centre Sonata Bis grant 2019/34/E/ST3/00405. VV was supported by the Complex Dynamics and Systems Program of the Army Research Office under grant W911NF-19-1-0268 and the Simons Foundation. This work was partially supported by the University of Chicago Materials Research Science and Engineering Center, which is funded by National Science Foundation under award number DMR-2011854.  DB and PS acknowledge Suropriya Saha and Piotr Witkowski for discussions.}
\bibliography{Bibliography.bib}

\end{document}